\documentclass[preprint, aps, pre, preprintnumbers,amsmath,amssymb,showpacs]{revtex4}

\usepackage[final]{graphicx}  

\newcounter{saveeqn}

\begin{document}
\title{
Bifurcation of standing waves into a pair of oppositely travelling waves with oscillating amplitudes caused by three-mode interaction
}

 \author{A.~Pinter\footnote[1]{Electronic address: kontakt@alexander-pinter.de}, M.~L\"ucke, and Ch.~Hoffmann}
 \affiliation{Institut f\"ur Theoretische Physik, Universit\"at des Saarlandes,
 Postfach 151150, D-66041 Saarbr\"ucken, Germany}

 \date{\today}

\begin{abstract}
A novel flow state consisting of two oppositely travelling waves (TWs) with
oscillating amplitudes has been found in the counterrotating Taylor-Couette
system by full numerical simulations.
This structure bifurcates out of axially standing waves that are nonlinear
superpositions of left and right handed spiral vortex waves with equal
time-independent amplitudes. Beyond a critical driving the two spiral TW modes
start to oscillate in counterphase due to a Hopf bifurcation. The trigger for
this bifurcation is provided by a nonlinearly excited mode of different
symmetry than the spiral TWs. A three-mode coupled amplitude equation model
is presented that captures this bifurcation scenario. The mode-coupling between two symmetry degenerate critical modes
and a nonlinearly excited one that is contained in the model can be expected to occur in other structure forming systems
as well.
  
\end{abstract}

\pacs{47.20.Ky, 47.20.Lz, 47.54.-r, 47.32.-y}


\maketitle


Many nonlinear structure forming systems that are driven out of equilibrium
show a transition to traveling waves (TWs) as a result of an oscillatory
instability \cite{CH93}. In the presence of spatial inversion symmetry in
one or more directions also a standing wave (SW) solution bifurcates which
is a nonlinear superposition of the two symmetry degenerated, oppositely
propagating TWs with equal amplitudes. 

SWs and TWs have a common onset as a result of a primary Hopf bifurcation.
But at onset only one of them is stable \cite{GS87,DI84}. 
Furthermore, there are mixed patterns with {\em non-equal} amplitude
combinations of the degenerated TWs that arise, e.g., via secondary
bifurcations at larger driving. The variety with temporally constant,
non-equal TW amplitudes can provide a stability transferring connection
between TWs that, e.g., are stable at onset and SWs that become stable later on 
\cite{PLH06}. 

The variety in which the TW amplitudes oscillate in time is the subject of this paper. 
This solution bifurcates out of the SW via a Hopf bifurcation. To be concrete we
investigate wave structures consisting of spiral vortices in the annular 
gap between counter rotating concentric cylinders of the Taylor-Couette 
system \cite{T94,CI94}. To that end we have performed numerical simulations of the Navier-Stokes 
equations (NSE) to reveal the bifurcation properties as well as the spatiotemporal structure of 
the novel oscillating mixed wave states. In addition, we provide coupled three-mode amplitude 
equations that capture this bifurcation to explain the underlying mode-coupling mechanism. 
We are not aware that these states have been reported so far in the Taylor-Couette 
literature. Furthermore, one can expect that the mode-coupling mechanism between two symmetry degenerate critical 
modes and the nonlinearly excited one that is described by our coupled amplitude equations and that drives the oscillatory 
instability is operating in other pattern forming systems as well.


The waves are realized by left handed spiral vortex (L-SPI) and right
handed spiral vortex (R-SPI) structures that are mirror images of each
other. The azimuthal advection by the basic circular Couette flow (CCF)
rotates 
both like rigid objects into the same direction as the inner
cylinder \cite{HLP04}. As a result of the enforced rotation the phases
of L-SPI and R-SPI travel axially into opposite directions. This system
offers an easy experimental and numerical access to forward bifurcating TWs
and SWs that are called ribbons (RIBs) \cite{RIBs} in the Taylor-Couette
literature. Being a nonlinear superposition of L-SPI and R-SPI the RIB
structure also rotates azimuthally, however, such that its oscillations in
axial direction form a SW.

Here we elucidate how such stable SWs loose stability 
to an oscillating state via a Hopf bifurcation. Therein, the interaction
with another nonlinearly excited, non-traveling mode induces the TW
constituents of the SW, i.e.,
the L-SPI and the R-SPI to oscillate in counterphase around a common mean. 
These oscillating mixed wave states that we call oscillating cross spirals
(O-CR-SPI) are quite robust. Thus, they should easily be observable in experiments. 

All these spiral structures are axially and azimuthally periodic. We have
focussed our simulations on patterns with axial wavelength $\lambda=1.3$
measured in units of the gapwidth and azimuthal wave number $M=2$. 
The numerical solutions of the NSE were obtained for a system with radius ratio 
$\eta=1/2$ by methods described in \cite{HLP04}.

\begin{figure}
\includegraphics[clip=true,width=8.6cm,angle=0]{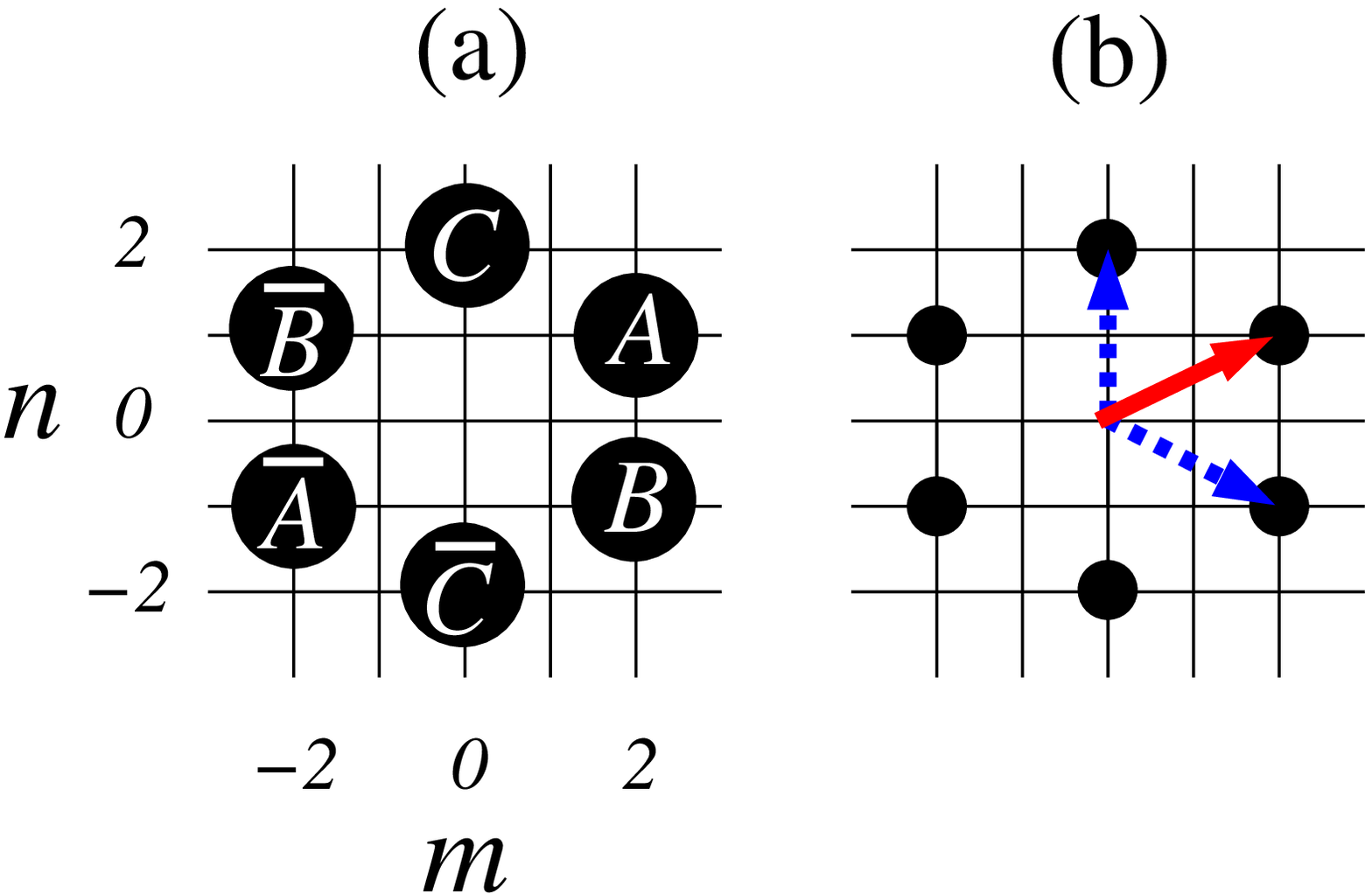}
\caption{(Color online) (a) Dominant modes and their complex conjugates in
the Fourier space of Eq. (\ref{modenansatz}). (b) Bilinear coupling of
modes $B$ and $C$ (dashed arrows) that drive oscillations of mode $A$
(solid arrow).
\label{Antrieb-O-CR-SPI}}
\end{figure}

{\it Control- and order parameters} --
The rotational velocities of the inner and outer cylinders are measured by
the respective Reynolds numbers $R_1$ and $R_2$. We fix $R_1=240$ \cite{phasediagram} 
and we introduce the reduced distance $\mu=(R_2-R_2^0)/|R_2^0|$ from the common
onset of SPI and RIB flow at $R_2^0=-605.5$ as control parameter. We
characterize the spatiotemporal properties of the vortex waves using the
Fourier decomposition
\begin{equation} \label{modenansatz}
f(r,\varphi,z,t) = \sum_{m,n} f_{m,n}(r,t)\,e^{i(m\varphi + nkz)}
\end{equation}
in azimuthal and axial direction. Here one has $f_{-m,-n}=\overline{f_{m,n}}$
with the overbar denoting complex conjugation. Order parameters are the moduli
$|A|,|B|,|C|$ and the time derivatives
$\dot\theta_A,\dot\theta_B,\dot\theta_C$ of the phases of the dominant modes
in the decomposition (\ref{modenansatz}) of, say, the radial velocity $u$ at
midgap $u_{2,1}=A=|A|e^{-i\theta_A}, u_{2,-1}=B=|B|e^{-i\theta_B}$, and
$u_{0,2}=C=|C|e^{-i\theta_C}$. Here, 
$A$ and $B$ are the amplitudes of the marginal L- and R-SPI modes. When both
are finite as, e.g., in the SW of the RIB state their nonlinear coupling
generates the $m=0$ $C$-mode below its threshold for linear growth: pure
$m=0$ stationary Taylor vortices bifurcate out of the CCF only later on.
Although $|C|$ itself remains small compared to $|A|, |B|$ in the RIB state its feedback on $A, B$ triggers the Hopf bifurcation of the O-CR-SPI: the
oscillations of, say, $|A|$ are driven by bilinear mode couplings of $BC$ as indicated in Fig.~\ref{Antrieb-O-CR-SPI}.

We also use the combined order parameters
\begin{equation}\label{EQ:def-comb}
S=\frac{|A|^2+|B|^2}{2}, \,  D=\frac{|A|^2-|B|^2}{2}, \, \Phi=\theta_C+\theta_B-\theta_A-\pi 
\end{equation}
that are better suited to describe the bifurcation of the O-CR-SPI
with oscillating $D(t)$ and $\Phi(t)$ out of the RIB state, $D=0=\Phi$.

\begin{figure}
\includegraphics[clip=true,width=8.6cm,angle=0]{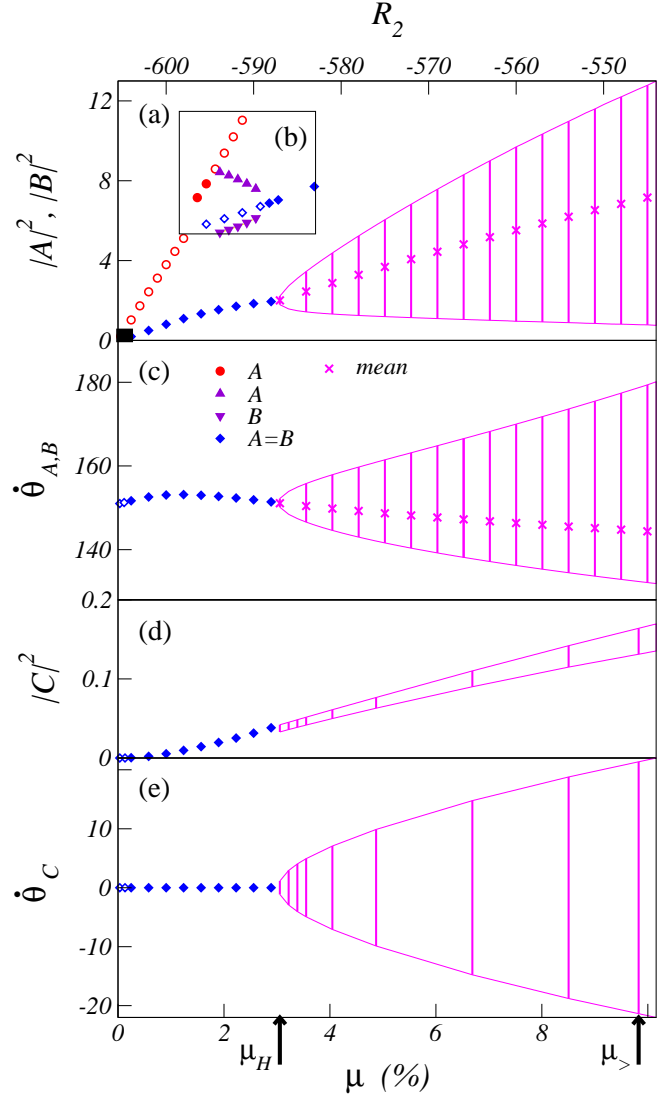}
\caption{(Color online) Bifurcation diagrams of SPI (red circles), RIB
(blue diamonds), CR-SPI (purple triangles), and O-CR-SPI (mangenta lines
and crosses) obtained from numerical simulations of the NSE versus $\mu$ and
$R_2$. SPI and CR-SPI are displayed only in (a) and (b); the latter shows
the blow-up of the rectangle near the origin of (a). Shown are the squared
mode amplitudes $|A|^2, |B|^2$ (a), (b) and phase velocities $\dot
\theta_A, \dot \theta_B$ (c) of the marginal modes and the same for the
nonlinear excited mode $C$ in (d) and (e). Filled (open) symbols denote
stable (unstable) solutions with time-independent amplitudes. Crosses refer
to temporal averages of the O-CR-SPI. Upper (lower) line shows the maximum
(minimum) of the oscillation range indicated by vertical lines. The arrow
at $\mu_H$ ($R_2=-587$) marks the Hopf bifurcation of the modulated SWs.
The second arrow at $\mu=\mu_>$ ($R_2=-546$) is inserted for later
reference.
\label{Bifurkdiagr}}
\end{figure}

{\it Bifurcation sequence} --
The pure TW shown by circles in Fig.~\ref{Bifurkdiagr} and
the SW solution $(A=B,C \neq 0)$ marked by diamonds bifurcate at $\mu=0$ out
of the unstructured CCF. Initially, the
SPI is stable and the RIB is unstable. But then there appears a stable 
cross-spiral (CR-SPI)
solution [triangles in Fig.~\ref{Bifurkdiagr}(b)]
which transfers stability from the SPI to the RIB. The moduli and phase 
velocities of these three structures are time-independent. 
At $\mu_H$ in Fig.~\ref{Bifurkdiagr} the RIB lose stability in a
supercritical Hopf bifurcation to the novel modulated state of O-CR-SPI. 
Increasing $\mu$ further beyond the range shown in Fig.~\ref{Bifurkdiagr} the
O-CR-SPI loses stability at $R_2 \approx -543$ to oscillating structures with
azimuthal wave number $M=1$ that are not discussed here.

\begin{figure}
\includegraphics[clip=true,width=8.6cm,angle=0]{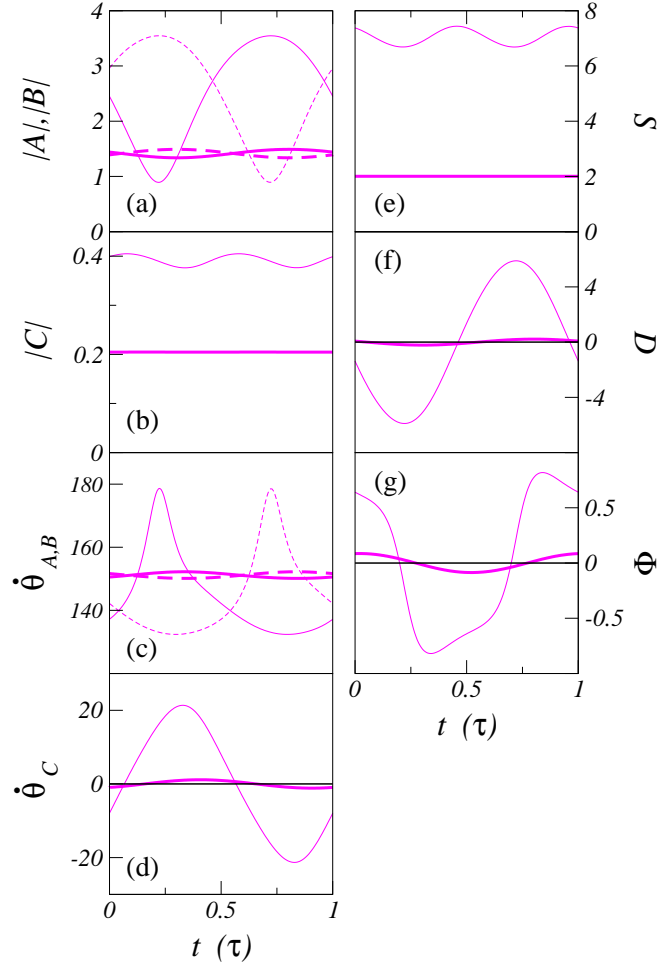}
\caption{(Color online) Time variation of O-CR-SPI over one modulation period $\tau$.
The left column shows moduli and phase velocities of $A$, $B$, and $C$.
In (a) and (c) solid lines refer to $A$ and dashed ones to $B$. The right
column contains the order parameters $S, D$, and 
$\Phi$ (\ref{EQ:def-comb}). Thick lines are modulation profiles close
to the Hopf threshold $\mu_H$ and thin ones those at the larger $\mu_{>}$ identified by second arrow in Fig.~\ref{Bifurkdiagr}.
\label{t-charakter.-Groessen}}
\end{figure}

\begin{figure}
\includegraphics[clip=true,width=8.6cm,angle=0]{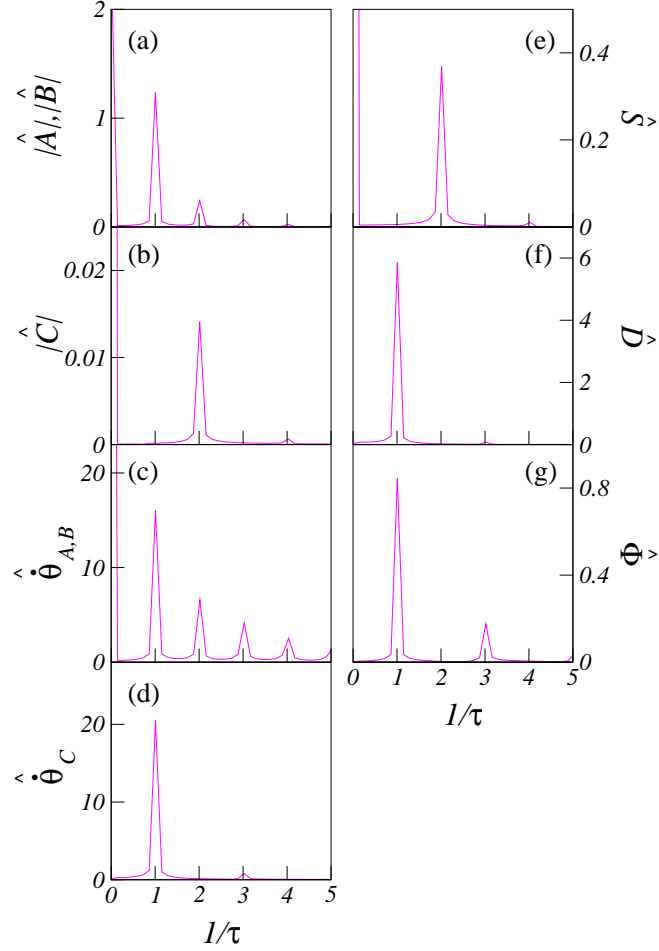}
\caption{(Color online) Fourier spectra of the modulation profiles shown by thin lines in
Fig.~\ref{t-charakter.-Groessen} for $\mu=\mu_>$ (cf. arrow in
Fig.~\ref{Bifurkdiagr}). Note that $|C|$ and $S$ oscillates with twice the frequency of the other quantities and that the spectra of $|C|$, $S$, and $D$
practically do not contain higher harmonics. The spectra of $\Phi$, $D$, and  $\dot \theta_C$ ($|C|$ and $S$) contain only peaks at $(2l+1)/\tau$ ($2l/\tau$) with $l=0,1,2\ldots$.
\label{fourier-charakter.-Groessen}}
\end{figure}

{\it Dynamics of the modulated SW} --
Figure~\ref{t-charakter.-Groessen} shows the temporal variation of
characteristic quantities of the O-CR-SPI over one modulation period $\tau$.
Thick lines refer to
$\mu$ immediately above onset $\mu_H$. Thin lines show behavior at a larger
value $\mu_>$ (arrow in Fig.~\ref{Bifurkdiagr}) that is close
to the end of the existence interval of O-CR-SPI. 
The moduli $|A|,|B|$ in Fig.~\ref{t-charakter.-Groessen}(a) and the phase velocities $\dot \theta_A, \dot \theta_B$ in Fig.~\ref{t-charakter.-Groessen}(c)
oscillate each in counter phase around a respective common mean. 
Also $\dot \theta_C$ oscillates. Furthermore, 
$|C|$ and the combined order parameter $S$ show small amplitude oscillations with twice the frequency of the other quantities. Close to onset all oscillations are harmonic with $|C|$ and
$S$ being practically constant. But at $\mu_>$ the oscillations of $|A|, |B|$
and $\dot \theta_A, \dot \theta_B, \Phi$ have become quite anharmonic, whereas
those of $S$, $D$, $|C|$, and $\dot\theta_C$ are still harmonic. The Fourier
spectra in Figs.~\ref{fourier-charakter.-Groessen}(a)-(g) of the temporal
profiles shown by thin lines in Figs.~\ref{t-charakter.-Groessen}(a)-(g)
reflect this behavior at $\mu_>$.

\begin{figure}
\includegraphics[clip=true,width=8.6cm,angle=0]{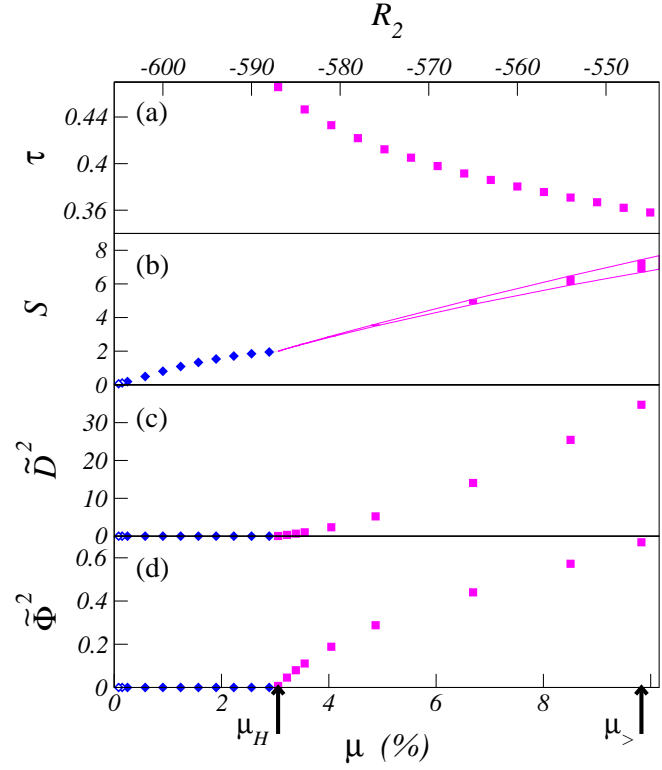}
\caption{(Color online) Bifurcation properties of RIB (blue diamonds) and O-CR-SPI (magenta lines and squares) obtained from numerical solutions of the NSE as functions of $\mu$ and $R_2$: (a) oscillation period $\tau$ of the modulation, say, of the moduli $|A|$ and $|B|$ of the O-CR-SPI, (b) S (thin lines delimit the oscillation range indicated by vertical bars), (c) and (d) squared oscillation amplitudes $\widetilde D$ of $D$  and $\widetilde \Phi$ of $\Phi$, respectively.  \label{Fig-Aufspalten}}
\end{figure}

In the RIB state the phases are such that $\theta_C(t)+\theta_B(t)-\theta_A(t)=\pi$. But as a consequence of the Hopf bifurcation $D$ as well as $\Phi$ oscillate in the O-CR-SPI. The squares of their oscillation amplitudes, $\widetilde D^2$ and $\widetilde \Phi^2$, increase at onset linearly with $\mu$ with a subsequent quadratic correction, cf. Figs.~\ref{Fig-Aufspalten}(c)-(d).
The monotonous decrease of the modulation period $\tau$ is shown in Fig.~\ref{Fig-Aufspalten}(a).
Note that the modulation amplitudes of $S$ in Fig.~\ref{Fig-Aufspalten}(b) and also of $|C|$ remain very small compared to those of $D$ and $\Phi$.

{\it Amplitude equations} --
The Hopf bifurcation behavior and the dynamics close to the transition from RIB to O-CR-SPI can be explained and described within a three-mode amplitude-equation approach. It reveals (i) how the rotationally symmetric $C$-mode is generated nonlinearly via the interaction of $A$ and $B$, i.e., of the M=2 SPI constituents in the RIB and (ii) how then $C$ -- after it has reached a critical size beyond $\mu_H$ -- induces amplitude oscillations in $A$ and $B$.

Invariance under axial translation and reflection of the Taylor-Couette system 
\cite{CI94} restricts the form of the three coupled amplitude equations to
\begin{subequations} \label{EQ-gekoppelte-AG-A-B}
\begin{eqnarray}
\dot A&=&A\ G\left(|A|^2,|B|^2,|C|^2 \right)+i\kappa{BC},\label{EQ-gek-1}\\
\dot B&=&B\ \widehat G\left(|A|^2,|B|^2,|C|^2 \right)+i\kappa{A\overline{C}},\label{EQ-gek-2}\\
\dot C&=&C\ H\left(|A|^2,|B|^2,|C|^2 \right)+\kappa_0 {A\overline{B}}.\label{EQ-gek-3}
\label{gekoppelte-GLE-O-CR-SPI}
\end{eqnarray}
\end{subequations}
With $\widehat G\left(|A|^2,|B|^2,|C|^2 \right)=G\left(|B|^2,|A|^2,|C|^2
\right)$ and $H(|A|^2,|B|^2,|C|^2)=\overline H(|B|^2,|A|^2,|C|^2)$
the equations are invariant under the operation $(A,B,C)
\leftrightarrow (B,A,\overline{C})$ which reflects the axial inversion
symmetry $z \leftrightarrow -z$. The functions
$G=G^{'}+iG^{''}$ and $H=H^{'}+iH^{''}$ are complex. The superscripts
$^{'}$ and $^{''}$ identify the real and imaginary parts, respectively.
The coupling constants $\kappa$ and $\kappa_0$ are real. 

Since only invariance under translation and reflection along one 
spatial direction has been used in deriving Eqs. (\ref{EQ-gekoppelte-AG-A-B}) 
our description of the phenomenon of a SW with oscillating TW components in terms of
Eqs. (\ref{EQ-gekoppelte-AG-A-B}) potentially 
applies to all bifurcating systems with O(2) symmetry in the center manifold, which
is quite common.

In the following we discard the coupling term $\kappa_0 A {\overline B}$. 
It is small in our case and, more importantly, we checked that it is not relevant 
for driving the Hopf oscillations.
They are generated by the coupling terms in (\ref{EQ-gek-1}) and (\ref{EQ-gek-2}) as 
we shall show in the next section.

The mechanism causing the Hopf bifurcation into the modulated SW can be better
isolated by rewriting the amplitude equations
(\ref{EQ-gekoppelte-AG-A-B}) in terms of the combined order
parameters (\ref{EQ:def-comb})
\begin{subequations}\label{EQ-gekoppelte-AG-S-D}
\begin{eqnarray}
\dot S &=& 2[D\ G_{-}^{'}+S\ G_{+}^{'}],\qquad \dot {|C|} =|C|\ H^{'},\label{EQ-gek-4}\label{EQ-gek-6} \qquad \\
\dot D &=& 2[S\ G_{-}^{'}+D\ G_{+}^{'}]-2\kappa|C| S^{\ast}\sin\Phi, \label{EQ-gek-5}\\
\dot \Phi &=& 2G_{-}^{''}+2\kappa\frac{|C|}{S^{\ast}} D \cos\Phi - H^{''} \label{EQ-gek-7},
\end{eqnarray}
\end{subequations}
where $S^{\ast}=S\sqrt{1-(D/S)^2} \simeq S$. Here we defined $G_{\pm}=(G \pm\widehat G)/2$. Note that $G_{+}$ and $H^{'}$ ($G_{-}$ and $H^{''}$) are even (odd) in $D$ \cite{explanation} as a result of the inversion symmetry. Hence, eqs. (\ref{EQ-gek-4}) are even in $D$. This in turn explains that $S$ and $|C|$ oscillate with twice the frequency of the other quantities in Fig.~\ref{t-charakter.-Groessen}. On the other hand, Eq. (\ref{EQ-gek-7}) is odd in $D$ and causes the absence of a peak at $2/\tau$ in Fig.~\ref{fourier-charakter.-Groessen}(g).

We have determined the specific functions $G$ and $H$ for our
specific system via fits to the numerically obtained bifurcation
branches of SPI, RIB, and CR-SPI in Fig. \ref{Bifurkdiagr} and to the pure Taylor 
vortex solution (not shown here) with $A=0=B, C \ne 0$, and half the spiral
wavelength. This produces the bifurcation behavior close to the Hopf threshold well. 
Note, however, that the Hopf bifurcation is a universal
phenomena of systems like (\ref{EQ-gekoppelte-AG-A-B},\ref{EQ-gekoppelte-AG-S-D}) 
that is not specific to the
Taylor-Couette system. This is most easily understood with the
help of the universal small-$D$ expansion of (\ref{EQ-gekoppelte-AG-S-D}) that 
results from the symmetry properties.

{\it Hopf bifurcation} --
For small $D$, i.e., close to the Hopf bifurcation threshold we can use the expansions
\begin{subequations}\label{EQ:gekoppelte-AG-S-D-small-D}
\begin{eqnarray}
G_{+}=G_{+}^{(0)}+\mathcal{O}\left(D^2\right), \quad
G_{-}=G_{-}^{(1)} D+\mathcal{O}\left(D^3\right),\\
H^{'}=H^{'(0)}+\mathcal{O}\left(D^2\right), \quad
H^{''}=H^{''(1)}D+\mathcal{O}\left(D^3\right).
\end{eqnarray}
\end{subequations}
Here the leading order terms $G_{+}^{(0)}, G_{-}^{(1)}, H^{'(0)}, H^{''(1)}$
still depend on $S$ and $|C|^2$. Inserting (\ref{EQ:gekoppelte-AG-S-D-small-D}) into (\ref{EQ-gekoppelte-AG-S-D})
and using the smallness of $\Phi$ yields a simplified model that is linear in $D$
\begin{subequations} \label{EQ:gekoppelte-AG-S-D-close-hopf}
\begin{eqnarray}
\dot S &=& 2S\ G_{+}^{'(0)},\qquad \dot {|C|} =|C|\ H^{'(0)},\label{EQ:gekoppelte-AG-S-D-close-hopf-1}\label{EQ:gekoppelte-AG-S-D-close-hopf-3}\\
\dot D &=& 2\left[S\ G_{-}^{'(1)}+\ G_{+}^{'(0)}\right]D - 2\kappa|C|S\Phi,\label{EQ:gekoppelte-AG-S-D-close-hopf-2}\\
\dot \Phi &=& \left[2G_{-}^{''(1)}+2\kappa\frac{|C|}{S} - H^{''(1)}\right]D .
\label{EQ:gekoppelte-AG-S-D-close-hopf-4}
\end{eqnarray}
\end{subequations}
It explains the Hopf bifurcation out of the RIB state and the 
O-CR-SPI properties close to onset. For example, $S$ and $|C|$ are
virtually constant because they are decoupled from $D$ and
$\Phi$ in the model eqs. (\ref{EQ:gekoppelte-AG-S-D-close-hopf}). 
Furthermore, Eq.~(\ref{EQ:gekoppelte-AG-S-D-close-hopf-4}) shows that 
$\Phi$ is enslaved by  $D$ and that the phase shift between them is $\tau/4$
as to be seen in Fig. \ref{t-charakter.-Groessen} close to $\mu_H$. This
justifies the solution ansatz
\begin{equation}\label{EQ-D-Phi}
D(t)=\widetilde D \cos(\omega_H t), \quad 
\Phi(t)=\widetilde \Phi \sin(\omega_H t)
\end{equation} 
where $\omega_H$ is the Hopf frequency. 

The latter is identified together with the bifurcation threshold $\mu_H$
by a linear stability analysis of the RIB fixed point
$D=0=\Phi, S=S_{RIB}(\mu), C=C_{RIB}(\mu)$ for which $ G_{+}^{'(0)}=0$
according to Eq.~(\ref{EQ:gekoppelte-AG-S-D-close-hopf-1}). Thus, the linearized equations for the stability-relevant deviations from this fixed point read
$\dot D =aD + b\Phi, \quad  \dot \Phi = c D$,
with coefficients $a=2SG_{-}^{'(1)}, b=-2\kappa|C|S,
c=2G_{-}^{''(1)}+2\kappa\frac{|C|}{S} - H^{''(1)}$ to be taken at the RIB fixed
point. Consequently, the location of the zero in $a(\mu)$ determines $\mu_H$
and the imaginary part of the eigenvalue at $\mu_H$, i.e., the Hopf
frequency is then given by $\omega_H^2=-bc\propto \kappa +h.o.t$, revealing that the coupling terms in (\ref{EQ-gek-1}) and (\ref{EQ-gek-2})
cause the Hopf bifurcation. Furthermore, $a(\mu)=\alpha
(\mu-\mu_H)$ with positive $\alpha$ to ensure decay of oscillations below
$\mu_H$ and growth above it.  

{\it Conclusion} --
The bifurcation of a novel spiral vortex structure with oscillating TW
amplitudes out of an SW is shown to be triggered by the coupling to a
nonlinearly excited mode when the latter exceeds a critical strength.
Since this novel state, in which the TW amplitudes oscillate in counterphase
around a common mean occurs quite robustly in a relatively wide parameter range it should be easily accessible to experiments.

Our results have been obtained by full numerical simulations and 
explained and confirmed by a coupled amplitude equation model
that captures the mode-coupling between two symmetry degenerate critical
modes and a nonlinearly excited one. Our bifurcation scenario can occur in
all systems with an O(2) symmetric center manifold, arising for example in
systems with translation and inversion symmetry, which is a quite general one. 
It has therefor the potential to occur also
in other structure forming systems, say, in hydrodynamics, chemical
reactions, or biological systems etc. where any 
two symmetry degenerate basic modes $A$ and $B$ couple similarly to a third
one, $C$, that is nonlinearly excited by them and that destroys the $A=B$
state once $C$ has reached a critical size.

This work was supported by the Deutsche Forschungsgemeinschaft.


\newpage

\begin{thebibliography}{999}
\bibitem{RIBs}
Stable RIBs were found \cite{TESM89} in a long system of
aspect ratio $\Gamma=36$ with radius ratio $\eta=0.727$ where theory
\cite{DI84,CI94} and numerical simulations \cite{TESM89} predicted a subcritical
transition. RIBs in end-plate-dominated short
systems ($\Gamma < 10, \eta=1/2$) were recently reported  \cite{LPA03-04} to come
with two different symmetries \cite{KP92}.
\bibitem{phasediagram}
A phase diagram of SPI, RIB, CR-SPI, and O-CR-SPI for $\lambda=1.3$ 
and an investigation of the wave number dependence of the bifurcation
porperties is provided in \cite{PLH08-2}.
\end{thebibliography}
\end{document}